\documentclass[prl,twocolumn,superscriptaddress,showpacs,floats,floatfix]{revtex4}
\usepackage{epsfig,dcolumn,amsmath,latexsym}
\begin{document}

\title{Metal contacts in carbon nanotube field effect transistors: Beyond the Schottky barrier
paradigm}

\preprint{1}

\author{J.\ J.\ Palacios}
\affiliation{Departamento de F\'\i sica Aplicada, Universidad
de Alicante, Campus de San Vicente del Raspeig, E-03690 Alicante, Spain.}

\author{P. Tarakeshwar}
\author{Dae M. Kim}
\affiliation{School of Computational Sciences, Korea Institute for Advanced Study,
207-43 Cheongnyangni-2-dong, Dongdaemun-gu, Seoul, South Korea.}

\date{\today}

\begin{abstract}
The observed performances of carbon nanotube field effect transistors are examined using first-principles quantum
transport calculations. 
We focus on the nature and role of the electrical contact of Au and Pd electrodes to open-ended semiconducting nanotubes,
allowing the chemical contact at the surface to fully develop through large-scale relaxation of the 
contacting atomic configuration.  We present the first direct numerical evidence
of Pd contacts exhibiting perfect transparency for hole injection as opposed to that of Au contacts. 
Their respective Schottky barrier heights, on the other hand, turn out to be
fairly similar for realistic contact models. 
These findings are in general agreement with experimental data reported to date, and show that
a Schottky contact is not merely a passive ohmic contact but actively influences the device I-V behavior. 
\end{abstract}


\maketitle
The superb performances of carbon nanotube (CNT) field effect transistors have been demonstrated over the last decade\cite{Tans:nature:98,Martel:apl:98,Martel:prl:01,Derycke:nl:01,Heinze:prl:02,Appenzeller:prl:02,Javey:apl:02,Wind:apl:02}. 
However, a number of factors determining the current voltage (I-V) characteristics have to be 
clarified for extensive device applications. A key factor in this context is the Schottky barrier (SB) formed at the 
interface between the source and drain metallic electrodes and the CNT. 
Traditionally, Schottky contacts have been introduced to serve as passive ohmic contacts. 
However, Schottky contacts in 
CNT field effect transistors play an active role in affecting the transistor action. 
For example, the drastic disparity of reported performances of CNT transistors has generally been attributed to the 
difficulty in controling the position of the Fermi energy, $E_{\rm F}$, with respect to the valence and conduction 
bands of the CNTs, when they are brought into contact to metal electrodes via different fabrication 
processes\cite{Derycke:apl:02,Cui:nl:03}. 
The different $E_{\rm F}$ locations should in turn give rise to different SB's and hence different 
I-V behavior. 

In the simplest Mott and/or Schottky picture for metal-semiconductor interfaces, the potential barrier of electrons 
is dictated primarily by the difference between the metal $E_{\rm F}$ and the CNT electron affinity. 
Accordingly, the gap of the semiconducting CNT, which is roughly inversely proportional to its 
diameter, is an important factor for determining the barrier for holes. 
Recently, a correlation has been shown between the diameter of the 
CNT and the on-current for negative gate voltage
of the device fabricated therein. Specifically the on-current increases with increasing 
CNT diameter\cite{Chen:nl:05}. Also, the work function of the metal electrodes has been shown to affect the 
device performance in a number of interesting ways. For instance, a metal electrode with a large work function, 
e.g. Pd, is shown to induce large on-currents for holes\cite{Javey:nature:03}, 
while metal electrodes having small work functions, 
e.g. Al, enhance the on-current of electrons\cite{Yang:apl:05}. When $E_{\rm F}$ lies near the midgap for intermediate 
values of the work function, the device is shown to exhibit an ambipolar 
behavior\cite{Yang:apl:05,Martel:prl:01,Javey:apl:02}. 

Early theoretical work discussed the apparent validity of the Mott-Schottky picture in CNT transistors\cite{Leonard:prl:00}.
However, this simple picture cannot account for the general features of observed I-V characteristics: 
(i) why the on-current does not scale exponentially with the work function in large band gap CNT's\cite{Chen:nl:05}, (ii) 
why the metal electrodes having similar work function induce different I-V behavior. Specifically, Pd and Au have the 
same work function, but in Pd-contacted CNT's the hole on-current was in the $\mu$A range\cite{Javey:nature:03}, 
whereas that 
attained in Au contacted CNT was in the nA range\cite{Derycke:apl:02}. Additionally, a perfect transparency was reported in 
Pd contacted small gap CNT's\cite{Javey:nature:03} and was attributed to the formation of ohmic contact for 
hole transport\cite{Javey:nature:03}. 
However, this also raises an interesting question, namely why the same ohmic contact has not been achieved 
with Au, having the same work function as Pd or with Pt\cite{Mann:nl:03}, which has a larger work function for that matter.

We present in this work a first-principles evaluation of the transport properties of large band-gap end-contacted CNT's. 
We have focused on the role of the Schottky contact and show that it plays a more pervasive role in determining the 
I-V behavior than what the traditional theories have predicted. 
We considered two different contact models, one intentionally
taken simple and one more realistic where we allow the chemical
bonding at the surface to fully develop via the relaxation of
contacting atomic positions. In an earlier work, it was shown
that relaxation of the atomic positions of both the CNT and
metal electrodes leads to distinct differences in the electronic
structure of the contacting region of the CNT\cite{Tarakeshwar:jpcb:05}. 
We have chosen for investigation Pd and Au electrodes 
since these metals are often utilized for device fabrication and 
have the same work function, but 
induce drastically different device performances. The experimental data are elucidated, using the calculated 
transmission curves, density of states (DOS), and $E_{\rm F}$  location within the gap. In particular, the $E_{\rm F}$ 
location is shown to be determined by the metal-induced density of states
in the gap of the CNT, which accomodates most of the interface charge dipole, and by the intrinsic CNT end states, which
pin $E_{\rm F}$ close to the mid-gap in the more simple contact models. For the more realistic contacts studied
the location of $E_{\rm F}$ turns out to be fairly similar for Au and Pd. Interestingly, however,
the CNT contacted to Pd is highly transparent for hole injection, providing the maximum possible
conductance of $ 4e^2/h = 2G_0$ at energies close to the valence band edge, while this is not the case for Au.

\begin{figure}[ht]
\includegraphics[width=0.9\linewidth]{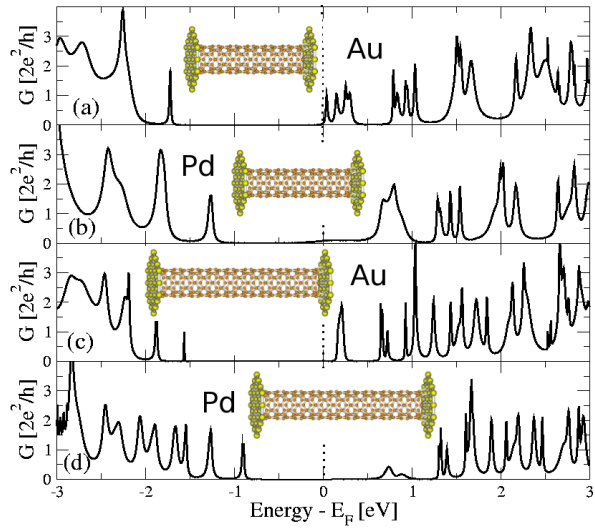}\\
\caption{(Color online). Conductance for an (8,0) CNT composed of 6 unit cells  contacted to Au (a) and Pd (b) as shown in 
the inset. Panels (c) and (d) show the same but for a 9 unit-cell CNT. \label{T_notemb} }
\end{figure}

\begin{figure}[ht]
\includegraphics[width=0.9\linewidth]{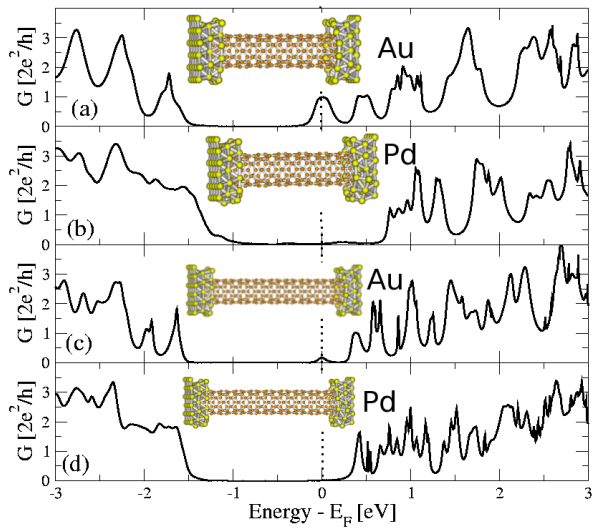}\\
\caption{(Color online). Conductance for a 6 [(a) and (b)] and a 9 [(c) and (d)] unit-cell (8,0) CNT contacted to Au 
[(a) and (c)] and Pd [(b) and (d)] electrodes
as shown in the inset. The nanotube ends are embedded in the metal one atom layer deep.\label{T_emb}}
\end{figure}

{\it Methodology and Results}.-- 
The transport calculations presented in this work are based on the quantum transport package 
ALACANT\cite{ALACANT:06}. The Kohn-Sham Green's function of the semiconducting CNT 
and a significant part of the electrodes (see insets in Figs. \ref{T_notemb} and \ref{T_emb}) 
is evaluated self-consistently with 
boundary conditions given by a parametrized description of the bulk electrodes implemented with the use of a 
Bethe lattice tight-binding model\cite{Palacios:prb:01,Palacios:prb:02}.
Other computational details are similar to those in Ref.\ \onlinecite{Palacios:prl:03}.

Figure \ref{T_notemb} shows the conductance of a (8,0) 6 unit-cell (UC) and of a 9 UC  
CNT with its open ends contacted to Au and Pd (111) layers as shown in the 
insets. To single out the influence of the electronic structure of the 
metal atoms from that of the contact geometry, we have considered an ideal and simple
contact model for both cases in which
electrode and CNT planes are kept at a fixed distance of 2.2 \AA $:$, i.e.
in an unoptimized electronic coupling configuration. 
The peaks in the conductance are the result of resonant transmission through quasi bound states associated 
with the discrete set of allowed
values of the longitudinal k-vector due to the finite length of the CNT.  Although not shown herein,
our calculations for infinite CNT's reveal two two-fold degenerate valence bands and a much larger
number of conduction bands in the energy range shown in this figure. This explains the dense transmission
spectrum above the gap and the sparse spectrum below where one can easily
establish a one-to-one correspondence between peaks and discrete states belonging to the valence band.

\begin{figure}[ht]
\includegraphics[width=0.9\linewidth]{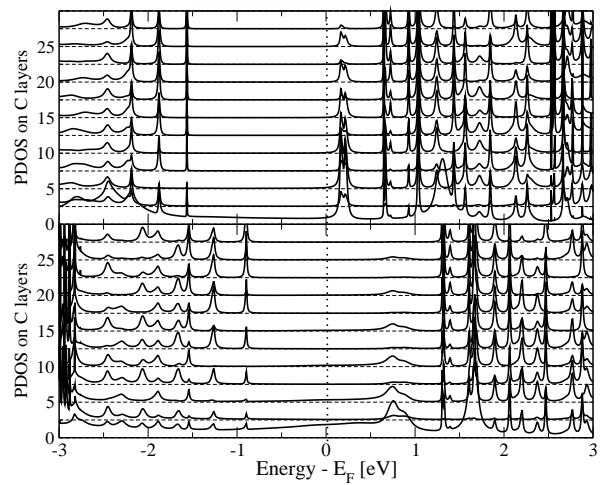}\\
\caption{Density of states projected on succesive carbon rings starting from 
the interface for the (c) (upper panel) and (d) (lower panel) cases shown in Fig.\ref{T_notemb}. Similar
features are obtained for (a) and (b).\label{DOS_9uc}}
\end{figure}

A closer look reveals a finite conductance in the gap for both metals.
The physical origin of this conductance can be clarified with the aid of Fig. \ref{DOS_9uc}, where the 
DOS projected on successive carbon rings from the interface is plotted for the 9 UC CNT. 
The DOS shows a broadened set of  peaks in the gap which decays into the CNT bulk, the decaying length being
longer for Au than for Pd. This
indicates an intrinsic origin in CNT surface or end states.
These states are commonly referred to as
metal induced gap states\cite{Leonard:prl:00}, but we would rather refer to them as intrinsic since they exist
regardless of the metal. 
In fact, calculations of finite CNT's (terminated by H atoms to simulate the bonding
to the surface) reveal 4 slowly decaying ($\approx 2 $nm) surface states in the gap.
These states are responsible for 
the finite conductance of short semiconducting CNT's at zero bias\cite{Pomorski:prb:04,Xue:prb:04:70:20,Rochefort:apl:01}.
The finite and nearly
constant DOS at the first interface C ring is apparently due to the chemical bonding between C and metal atoms. 
A fully developed energy gap is already visible from the third C ring on and no band bending can be 
seen\cite{Xue:prb:04:70:20}, at least at these length scales\cite{Leonard:prl:00}.
 
We note that $E_{\rm F}$ always lies below the lower edge of the surface state band for both 
Pd and Au contacts. The location of $E_{\rm F}$ is primarily determined by the contact-induced DOS at the first C layer
which accomodates most of the charge transferred (from the metal to the CNT in the cases studied). 
Within the inherent uncertainty to Mulliken population analysis, 
the charge transferred is essentially the same for both metals 
($\approx 4$ electrons for Au and $\approx 3.8$ for Pd),
but the the metal-induced DOS is smaller for Au than for Pd (see Fig. \ref{DOS_9uc}). As a consequence,  
the Fermi level moves up within the gap in the former case  until it gets
pinned by surface states as shown in Figs. \ref{T_notemb}(a) and (c).
Since the surface state band arises from fairly extended surface states, if $E_{\rm F}$ lies above the band,
allowing thereby electrons occupying these gap states, the excess charge would render the interface chemical
bonding unstable\cite{Pomorski:prb:04}.
This clearly shows the consistency of our computational results of charge redistribution accompanying 
the metal contact. Interestingly, due to the differences in the DOS within the gap between Au and Pd,
the SB for these contact models turn out to be different,
despite the fact that Pd and Au possess practically the same work functions.

\begin{figure}[ht]
\includegraphics[width=\linewidth]{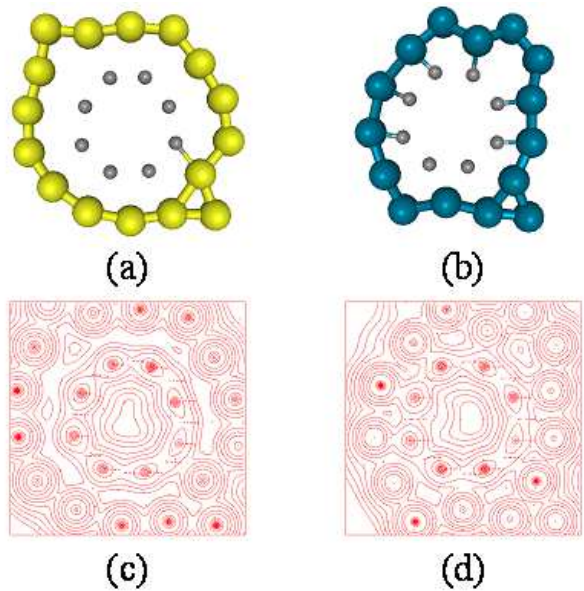}
\caption{(Color online). Atomic cross section of the metal-CNT interface  for  Au (a) and Pd (b) electrodes.
Contour density plot at the interface for Au (c) and Pd (d).}
\label{crosssection}
\end{figure}

Not only the DOS induced in the gap by the two metals is different. They also
induce a distinctly different transmission.  
The CNT-Pd system exhibits broader conduction peaks below the gap which  
indicates a stronger hybridization strength of Pd atoms at these energies.
All these results cannot be interpreted in light of conventional 
theory in which the properties of the contacting interface are mainly dictated by the metal work function. 
Rather it points to the importance of electronic coupling operative at the surface.  We thus
consider now a more realistic embedded contact geometry in which the end layers of the CNT are dipped 
into Pd and Au electrodes (see insets in Fig. \ref{T_emb}). 
(Experimental evidence that electrical contact only occurs at
the edge of the metal electrode\cite{Mann:nl:03} in part supports this contact model, although
in real devices the CNT is deeply buried into the electrode\cite{Zhu:nl:06}.)
In this case the surface atomic contact has been optimized, i.e., the atomic positions are 
allowed to relax to render the minimum energy and an optimized coupling. Note the drastic improvement of the transmission
especially in the Pd-contacted CNT's compared to that shown in
Fig. \ref{T_notemb}. The resonance transmission peaks appearing close to the valence band edge in 
Fig. \ref{T_notemb} are here practically fused into a broad band   and the 
conductance presents the maximum possible transmission of $4e^2/h$ [Figs. \ref{T_emb}(b) and (d)]. For the case of Au, 
the improvement is there but the effect is not as conspicuous as in Pd. These improved transport 
properties can be attributed to (i) increased contact area due to the embedded geometry, 
hence more extensive hybridization and (ii) the optimized atomic configurations enabling 
maximum chemical bonding at the surface. These effects will be further discussed in correlation with Fig. 
\ref{crosssection}.  Transmission in the conductions band is, on the other hand, not significantly affected by the relaxed
 atomic configurations. The location of  $E_{\rm F}$ in the gap is now fairly similar for
both Au and Pd electrodes and lies closer to the conduction edge due to a larger charge transfer.

Figures \ref{crosssection}(a) and (b) 
show the cross-sectional profiles of the optimized atomic configurations and the associated 
charge distribution [(c) and (d) panels] 
in the embedded layers for Au and Pd. Note the drastic differences in hybridization, 
specifically a strong chemical bonding for the case of Pd and a relatively weak bonding for Au. 
The surface atoms of gold exhibit a reduced $d$ delocalization and $sd$ hybridization\cite{Citrin:prl:78}. 
On the other hand, the enhanced $sd$ hybridization and the concomitant delocalization of the $d$ electrons 
ensure that the repulsion between the Pd and C atoms of the CNT is much smaller than that 
between Au and C atoms\cite{Mueller:prb:70,Louie:prl:78}. Consequently, the Pd atoms are more strongly bound to the  
face of the CNT\cite{Durgun:prl:03,Maiti:cpl:04}. This results in an enhanced charge redistribution at the Pd-C
interface than at the Au-C interface [see Figs. \ref{crosssection} (b),(d)]. The improved charge exchange at the 
 Pd-C interface also leads to an enhanced redistribution of the  electron density of 
the contacting layer of the CNT\cite{Tarakeshwar:jpcb:05}.
An additional effect of $sd$ hybridization in Pd is that its Fermi surface is almost entirely 
composed of $d$-like states\cite{Mueller:prb:70}, while that of Au is composed of $s$-like states, with $d$ DOS lying 
below $E_{\rm F}$ by about 2 eV\cite{Shirley:prb:72}.
The surface bonding is thus expected to
give rise to substantial changes in the electronic contact properties of a CNT due to the 
extended nature of hybridized states as confirmed evidently in Fig. \ref{T_emb}. 

{\em Summary and concluding remarks}.--– 
We have examined metal-contacted small-diameter CNT's in which case $E_{\rm F}$ lies mainly in the gap at the interface, 
as confirmed by experiments\cite{Chen:nl:05}. What our calculations show is that Pd
possess several attributes making it an excellent electrode for hole conduction, compared to Au.
We can conclude that, although the location of $E_{\rm F}$ in the gap is, in general, a key factor dictating 
the contact resistance, barrier potential, and the overall device performance,
the observed superior performance of the Pd-contacted p-channel CNT 
field effect transistor can unambiguously be
attributed to the strong electronic coupling at the surface facilitated by a relaxed atomic configuration.
On the contrary, a near ideal contact is never obtained for electron injection in either case. All this  
is consistent with generally reported p-type conduction in CNT field effect transistors.
For simple contact models the SB differs between Au and Pd, but the difference seems to  vanish
for more realistic contact models.  The precise location of $E_{\rm F}$ within the gap depends 
on the CNT surface DOS, which presents strong pinning properties, and on the highly localized
metal-induced DOS, with much weaker pinning attributes.  Both, in turn, strongly depend on the details of 
atomic structure of contact, making it difficult to extract universal conclusions. 

To conclude, Schottky contacts cannot be viewed merely as the passive ohmic contact. Rather it is an integral 
part of active elements of the device influencing the transistor action. 
Thus it is important that the detailed role of the 
contacts be further investigated, in correlation with the chemical nature of surface electronic coupling
and the sticking properties of metal atoms to the CNT. 
Additionally, the surface band bending should be analyzed together with the effect of both transverse 
and longitudinal electric field as induced by the gate and drain voltages. Finally, the device 
performance should be further investigated as a function of the diameter and length of the CNT.
These are crucial for understanding the operational principle of the device and formulating a compact I-V model.

{\it Acknowledgements}.--
J.J.P. acknowledges discussions with F. Leonard and J. Fern\'andez-Rossier.  
This work has been partially funded by Spanish MEC under grant  FIS2004-02356. 
P.T. and D.M.K. acknowledge the computational resources provided
under the KISTI strategic support program.

\end{document}